\documentclass[twocolumn,twoside,slac_two,floatfix]{revtex4}
\usepackage{graphicx}
\usepackage{fancyhdr}
\begin{document}

\title{Dependence of the optical brightness on the gamma and X-ray properties of
GRBs}

\author{L.G. Bal\'azs}
\affiliation{MTA CSFK Konkoly Observatory, Budapest, Hungary}
\affiliation{E\"otv\"os University, Budapest, Hungary}

\author{Z. Bagoly}
\affiliation{E\"otv\"os University, Budapest, Hungary}

\author{I. Horv\'ath}
\affiliation{National University of Public Service, Budapest,
Hungary}

\author{J. K\'obori}
\affiliation{E\"otv\"os University, Budapest, Hungary}

\author{D. Sz\'ecsi}
\affiliation{E\"otv\"os  University, Budapest, Hungary}

\author{A. M\'esz\'aros}
\affiliation{Charles University, Prague, Czech Republic}

\begin{abstract}
The Swift satellite made a real break through with measuring simultaneously the
gamma X-ray and optical data of GRBs, effectively.  Although, the satellite
measures the gamma, X-ray and optical properties almost in the same time a
significant fractions of GRBs remain undetected in the optical domain. In a
large number of cases only an upper bound is obtained. Survival analysis is a
tool for studying samples where a part of the cases has only an upper (lower)
limit. The obtained survival function may depend on some other variables. The
Cox regression is a way to study these dependencies. We studied the dependence
of the optical brightness (obtained by the UVOT) on the gamma and X-ray
properties, measured by the BAT and XRT on board of the Swift satellite. We
showed that the gamma peak flux has the greatest impact on the afterglow's
optical brightness while the gamma photon index and the X-ray flux do not. This
effect probably originates in the energetics of the jet launched from the
central engine of the GRB which triggers the afterglow.  

\end{abstract}

\maketitle

\thispagestyle{fancy}

\section{INTRODUCTION}
A significant achievement of the Swift satellite is the
simultaneous detection of the physical properties of the
gamma ray bursts in the gamma, X-ray and optical domain,
measured by the BAT, XRT and UVOT instruments on board of
the satellite.

Following the alert given by BAT the satellite starts to
slew and after reaching the position of the burst the XRT
and UVOT make measurements in the X-ray and optical domain,
respectively.

Although, a significant fraction of the bursts is detected
by the XRT as well, it is not the case with the UVOT where
at a remarkable fraction of the events only an upper limit
of the optical brightness is obtained.

From theoretical point of view the measured optical and
gamma properties may be given by completely different
phenomena, their observational relationship, if there is
any, would be an important constraint for the possible
models.  To study this relationship it would be a serious
bias if we take into account only those cases where all
properties, i.e. gamma, X-ray and optical, are measured.

Survival analysis is a way to make use the information which
is inherent in the value of the upper bound of the optical
brightness. Cox regression is a tool for studying the
dependence of the survival function (a result of the
analysis) on some background variables, the covariates
(gamma and X-ray properties in our case).

In the following we use Cox regression to study the
dependence of the distribution of the UVOT detected optical
brightness on measured gamma and X-ray properties.

\section{MATHEMATICAL SUMMARY}
Let we have a $t$ stochastic variable with $f(t)$
probability density. The $S(t)$ survival function is defined
by
\begin{equation}\label{sfunc}
\int\limits^t_{-\infty}f(t')dt'=F(t)=1-S(t)
\end{equation} 
where $F(t)$ means the probability distribution function.
Actually, the $S(t)$ survival function is its complement
$(F(t) + S(t) = 1)$. \cite{kam} showed that $S(t)$ can be
estimated bias free even in the case when some of the values
in the $t_1, t_2, \ldots, t_n$ observed sample are only
lower bounds (censored). The ratio of $f(t)$ to $S(t)$ is
called the hazard function:
\begin{equation}\label{hfunc}
h(t)=\frac{f(t)}{S(t)}=-\frac{S'(t)}{S(t)}=-\frac{d}{dt}log[S(t)]
\end{equation}
The $h(t)$ hazard function characterizes the risk that
in the $[t,\infty]$ range $(S(t)$ gives its probability) an
event will happen in the $[t,t+dt]$ interval (its
unconditional probability is $f(t)dt)$. The hazard function
may depend on background variables (the covariates). The Cox
model (\cite{cox}) assumes that this dependency can be
written in the form of
\begin{equation}\label{lhfunc}
log[h(t)]=\alpha (t)+B_1x_1+B_2x_2+\ldots+B_mx_m
\end{equation}
where $x_1,x_2,\ldots,x_m$ are the covariates while the
$\alpha(t)$ arbitrary function and the $B_1, B_2, \ldots,
B_m$ constants have to be determined during the procedure of
the Cox regression. If all these constants are equal to zero
the $\alpha(t)$ function is identical with the logarithmic
hazard function. The value of the constants characterize the
strengths of the influence of covariates on the hazard and,
consequently on the survival function.

\section{DESCRIPTION OF THE DATA}
We used for the present analysis the data available in the Swift
table\footnote{
http://swift.gsfc.nasa.gov./docs/swift/archive/grb\_table}
recorded until the date of 03/03/2012, in particular, the V
magnitude as a dependent variable, Duration, Fluence, Peak
flux, Photon index and early X-ray flux, as covariates in
the analysis.  Except of the Photon index we used
logarithmic values in order to suppress the impact of the
outliers on the regression.

Since the optical brightness of the GRB afterglow is
seriously dimmed by the foreground Galactic extinction we
excluded the cases with the latitude of 
$|b| < 15^o (|\sin b| < 0.26)$.

\begin{figure}
\includegraphics[width=65mm]{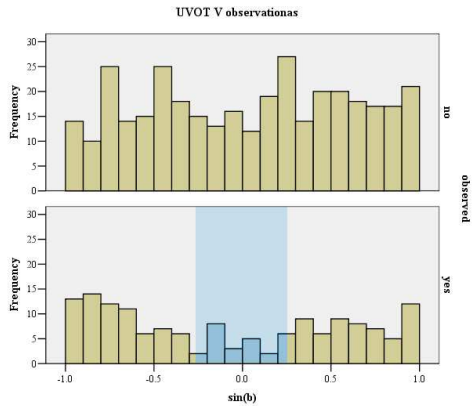}
\caption{Distribution of the GRB positions according to the
Galactic latitude. Note the depression in the distri\-bution
close to low latitudes due to the foreground extinction of
the Galactic dust. Shadowed region marks the excluded area.}
\label{sinb}
\end{figure}

In the case if no afterglow was observed a lower bound in
the stellar magnitude (upper bound for the observed
brightness) was obtained. 
One can infer from Fig.~\ref{sinb} that at low latitudes
the depression is not present in the distribution of the
cases where only a lower V magnitue bound was determined.

\section{COX REGRESSION}

Before we make the Cox regression running we have computed
the bivariate correlations between the variables included in
the analysis. In this procedure we have taken into account
only those cases where both variable used for computing the
correlation had measured values. To overcome the problem
with the outliers we computed the Spearman's rank
correlation which is not sensitive to these.

\begin{table}
  \centering
  \caption{Spearmann's correlations (numbers in bold face mark significant correlations)}\label{Spe}
  \begin{tabular}{|l l|r|r|r|r|r|r|}
    \hline
    \  & \  & {\bf T90} & {\bf Flu} & {\bf Peak} & {\bf Pind} & {\bf Xflu} & {\bf V} \\
    \hline
    {\bf T90} & corr.& 1.000 & {\bf .646} & -.066 & {\bf .155} & {\bf .408} & -.024 \\
    \  & sign. & - & $<$.001 & .149 & .001 & $<$.001 & .646 \\
    \hline
    {\bf Flu} & corr. & {\bf .646} & 1.000 & {\bf .539} & {\bf -.148} & {\bf .471} & {\bf -.204} \\
    \  & sign.& $<$.001 & - & $<$.001 & .001 & $<$.001 & .000\\
    \hline
    {\bf Peak} & corr. & -.066 & {\bf .539} & 1.000 & {\bf -.282} & {\bf .141} & {\bf -.289} \\
    \  & sign. & .149 & $<$.001 & - & $<$.001 & .018 & $<$.001 \\
    \hline
    {\bf Pind} & corr.  & {\bf .155} & {\bf -.148} & {\bf -.282} & 1.000 & .049 & .004 \\
    \  & sign. & .001 & .001 & $<$.001 & - & .406 & .944 \\
    \hline
    {\bf Xflu} & corr. & {\bf .408} & {\bf .471} & {\bf .141} & .049 & 1.000 & -.109\\\
    \  & sign. & $<$.001 & $<$.001 & .018 & .406 & - & .080 \\
    \hline
    {\bf V} & corr. & -.024 & {\bf -.204} & {\bf -.289} & .004 & -.109 & 1.000 \\
    \  & sign. & .646 & $<$.001 & $<$.001 & .944 & .080 & - \\
    \hline
  \end{tabular}
\end{table}

\begin{figure}
\includegraphics[width=65mm]{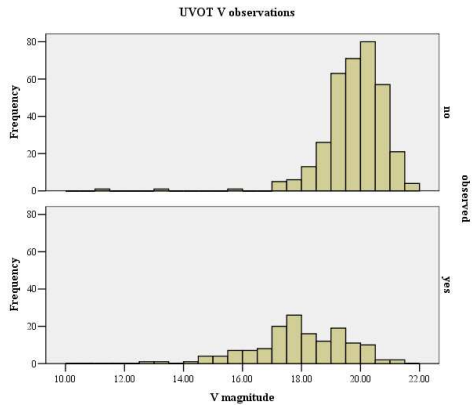}
\caption{Distribution of the afterglows' brightness in
visual magnitude. Note the increasing fraction of censored
data as we are approaching the detection limit.}
\label{vdistr}
\end{figure}

In Table~\ref{Spe} we used all cases having measured values
in both variables used in the analysis, pairwise. We marked
with bold face  where the correlation coefficients differ
significantly from zero. As we are approaching the detection
limit of UVOT, however, only a lower magnitude limit is
obtained  in an increasing number of cases (see
Fig.~\ref{vdistr}). Of course, cases having only an lower
limit in V have to be excluded from computing the
correlation of this variable with the other ones.

\begin{table}
  \centering
  \caption{Estimated $B$ coefficients in Eq.~(\ref{lhfunc}) (number in
	   bold face means that the corresponding $B$
	   differs from zero significantly.)}
  \label{cox}
  \begin{tabular}{|l|r|r|r|r|}
    \hline
    \  & {\bf B} & {\bf Wald} &\ \ \  {\bf df} & \ \  {\bf sign.} \\
    \hline
    {\bf logT90} & .732 & 6.346 & 1 & {\bf .012} \\
    \hline
    {\bf logFlu} & -.809 & 4.099 & 1 & {\bf .043} \\
    \hline
    {\bf logPeak} \ \  & \ 1.886 & \ 25.438 & 1 & {\bf .000} \\
    \hline
    {\bf Pind} & .058 & .055 & 1 & .815 \\
    \hline
    {\bf logXflu} & -.003 & .001 & 1 & .973 \\
    \hline
  \end{tabular}
\end{table}

Results of the Cox regression are summarized in
Table~\ref{cox}.  Seemingly, the duration, gamma fluence and
peak flux has a significant impact on the distribution of
the optical brightness, while the gamma photon index and the
early X-ray flux do not.

\section{CONCLUSIONS}
We performed Cox regression in order to look for the impact
of the gamma and X-ray properties of the GRBs on the
afterglows' optical brightness. This approach is necessary
since in a significant fraction of cases only an upper bound
of the optical brightness (lower bound in the V magnitude)
can be determined.

The analysis demonstrated that among the $B$ coefficients
in Eq.~(\ref{lhfunc}) belonging to the duration, fluence and
peak flux  differ significantly from zero.
Nevertheless, it is not the case with the gamma photon index
and the early X-ray flux.

The reason for the impact of some gamma properties on the
optical brightness is probably lying in the energetics of
the jet launched from the central engine of the GRB which
triggers the afterglow in the surrounding interstellar
matter.

\begin{acknowledgments}
This work was supported by OTKA grant K077795, by OTKA/NKTH
A08- 77719 and A08- 77815 grants (Z.B.), by the Grant Agency
of the Czech Republic grant P 209/10/0734 (A.M.), and by the
Research Program MSM0021620860 of the Ministry of Education
of the Czech Republic (A.M).
\end{acknowledgments}

\end{document}